\documentclass{IEEEtran4PSCC}
\ifCLASSINFOpdf
\else
\fi
\usepackage[cmex10]{amsmath}
\usepackage[all]{xy}
\usepackage{color}
\usepackage{multirow}
\usepackage{accents}
\usepackage{algorithm}
\usepackage{cite}
\usepackage{algpseudocode}
\usepackage{subcaption}
\usepackage{bm}
\usepackage{graphicx}
\usepackage{color}
\usepackage{wrapfig}
\usepackage{epsf}
\usepackage{times}
\usepackage{url}
\usepackage{amssymb, amsmath, mathtools, amsthm}
\usepackage{nomencl}
\usepackage{comment}
\usepackage{enumitem}
\makenomenclature
\newtheorem{theorem}{Theorem}

\newtheorem{lemma}{Lemma}
\newtheorem{assumption}{Assumption}
\definecolor{RED}{rgb}{0.6,0.,0.}
\definecolor{BLUE}{rgb}{0.,0.,0.6}
\definecolor{GREEN}{rgb}{0.,0.6,0.}
\definecolor{MALINA}{rgb}{0.6,0.,0.6}
\definecolor{YELLOW}{rgb}{0.8,0.8,0}
\newcommand{\squeezeup}{\vspace{-1.5 mm}}

\begin{document}
%
\title{Exact Topology and Parameter Estimation in Distribution Grids with Minimal Observability}

\author{
\IEEEauthorblockN{Sejun Park$^{*,\dagger}$}
\IEEEauthorblockA{
($*$) KAIST, Daejeon, Korea\\
\& ($\dagger$) LANL, 
Los Alamos,\\ New Mexico, USA\\
sejun.park@kaist.ac.kr}
\and
\IEEEauthorblockN{Deepjyoti Deka$^\dagger$}
\IEEEauthorblockA{
($\dagger$) LANL, Los Alamos,\\ New Mexico, USA\\ deepjyoti@lanl.gov}
\and
\IEEEauthorblockN{Michael Chertkov$^{\dagger,\mathsection}$}
\IEEEauthorblockA{
($\dagger$) LANL, Los Alamos,\\ New Mexico, USA\\
\& ($\mathsection$) Skoltech, Moscow, Russia\\
chertkov@lanl.gov}
}


\maketitle

\begin{abstract}
Limited presence of nodal and line meters in distribution grids hinders their optimal operation and participation in real-time markets. In particular lack of real-time information on the grid topology and infrequently calibrated line parameters (impedances) adversely affect the accuracy of any operational power flow control. This paper suggests a novel algorithm for learning the topology of distribution grid and estimating impedances of the operational lines with minimal observational requirements - it provably reconstructs topology and impedances using voltage and injection measured only at the terminal (end-user) nodes of the distribution grid. All other (intermediate) nodes in the network may be unobserved/hidden. Furthermore no additional input (e.g., number of grid nodes, historical information on injections at hidden nodes) is needed for the learning to succeed. Performance of the algorithm is illustrated in numerical experiments on the IEEE and custom power distribution models. 
\end{abstract}

\thanksto{This work is supported by the US DOE/OE's Grid Modernization Laboratory Consortium (GMLC) program.}
\section{Introduction}
\label{sec:intro}
Distribution grids include low and medium voltage lines that enable supply of power to end-users/ loads. With the advent of smart grids, new resources like controllable loads, small-scale/household renewable generators (e.g. solar panels) and storage units (e.g. batteries, electric vehicles) have been introduced in the distribution grid. This paradigm shifting change is turning traditionally passive and largely static distribution grids, that acted solely as power sinks, into dynamic, reconfigurable and active resources for novel controls. Emerging smart grid control technologies, relying on these dynamic and reconfigurable features, include participation in demand response, use of batteries in frequency regulation and inter-household energy settlements/transactions. These new technologies also require much more accurate estimation of the grid characteristics. The radial topology of current operational lines, and their impedances are the most significant of these frequently changing characteristics of the smart distribution grids. However, reliable and real-time estimation of the distribution grid topology and line impedances is impeded by limited observability. Even though Phasor Measurement Unit (PMU) technology has become available recently, it still largely limited to power transmission systems \cite{hoffman}. It is not obvious if a wide-spread (full coverage) use of the PMU technology will ever be economically justified at the distribution level. Moreover, access to underground distribution grid in urban areas (e.g. New York City) is technically challenging. These constraints/limitations justify importance of developing techniques capable of estimating operational topology and line impedances in the situation of infrequent calibration and sparse access. 

In spite of the access limitations 
PMUs, micro-PMUs \cite{micropmu}, FNETs \cite{FNET} have being placed into many distribution grids. In addition, and most importantly for the setting discussed in the manuscript, smart end-user measurement units, such as ones associated with smart house-hold devices and EVs, have been installed. These modern end-user devices have the ability to record and communicate nodal voltages and injections. Inspired by this novel capabilities we analyze in this manuscript the joint problem of topology and impedance estimation using measurements collected from smart meters located only at the end/terminal nodes of the smart distribution grids. 

\subsection{Prior Work}
Topology estimation in the power grid is an active area of research. Researchers have proposed different methods depending on availability and type of measurements. \cite{ramstanford} uses cycle basis and maximum likelihood tests to reconstruct topology from line measurements. For measurements of nodal voltages collected at all nodes, \cite{bolognani2013identification, dekairep} uses the signs of inverse covariance of complex voltages to identify the operational lines. In a similar regime, \cite{dekatcns,dekaecc} presents greedy schemes based on trends in second moments of voltage magnitudes to identify grid topology. \cite{deka2016estimating,ram_loop} 
utilizes statistical independence tests, in the setting of a graphical model associated with nodal voltages, to reconstruct topology. A number of data driven and model-free schemes, using signature and regression based methods, were developed to reconstruct topology and line parameters \cite{berkeley, arya,sandia1}. In the work \cite{dekasmartgridcomm}, most closely related to this manuscript, an iterative scheme to estimate topology and line impedances from voltage measurements is proposed. 

A common feature of the aforementioned papers is that they all rely on availability of nodal measurements (voltage and/or injection) at all nodes of the grid. Reconstruction in the case with limited voltage observability as discussed in \cite{dekatcns, dekaecc} require a minimum separation between observed nodes. Further, \cite{dekasmartgridcomm} proposed an iterative algorithm for topology estimation with missing nodes, however, it requires to know the all operational/non-operational line and their impedances. These assumptions can be easily broken due to non-ubiquitous presence of nodal meters and lack of historical and state-estimation information. 
In this manuscript, we overcome these impediments and suggest an efficient algorithm which estimates jointly operational topology and line impedances. 

\subsection{Technical Contribution}
We develop a provable method/algorithm for topology estimation in the radial distribution grids from samples of nodal voltage and injection measurements collected from smart meters at the end-user locations. The remaining nodes (including all intermediate nodes) in the grid may be unobserved, i.e. their states are unmeasured and further even the knowledge of their existence and number may be unknown. This represents a realistic scenario where majority of measurements are limited to residential devices installed at the end-user locations. Our reconstruction is model-based. We consider a linearized power flow model \cite{89BWa,89BWb,bolognani2016existence, dekatcns}. Under this model, we develop an algorithm which first learns the impedance distance between any two observed nodes in the radial grid. Using these reconstructed impedance distances, we utilize the recursive grouping algorithm \cite{choi2011learning} to learn the operational topology and associated line impedances. We demonstrate performance of the algorithm on some useful toy examples and then present results of experiments based on ac power flows on realistic IEEE test cases. To the best of our knowledge, this is the first work which provides guaranteed topology and impedance reconstruction in the distribution grid where only terminal nodes are observed. 

The rest of the manuscript is organized as follows. Section \ref{sec:preliminary} introduces nomenclature and power flow relations in the distribution grid. Our main algorithm is described in Section \ref{sec:main}. Numerical experiments on IEEE test cases are presented in Section \ref{sec:experiments}. Finally, Section \ref{sec:conclusions} is reserved for conclusions and discussion of future work.

\section{Distribution Grid and Power Flow Model}\label{sec:preliminary}
\begin{figure}[ht]
 \centering
\includegraphics[width=0.35\textwidth]{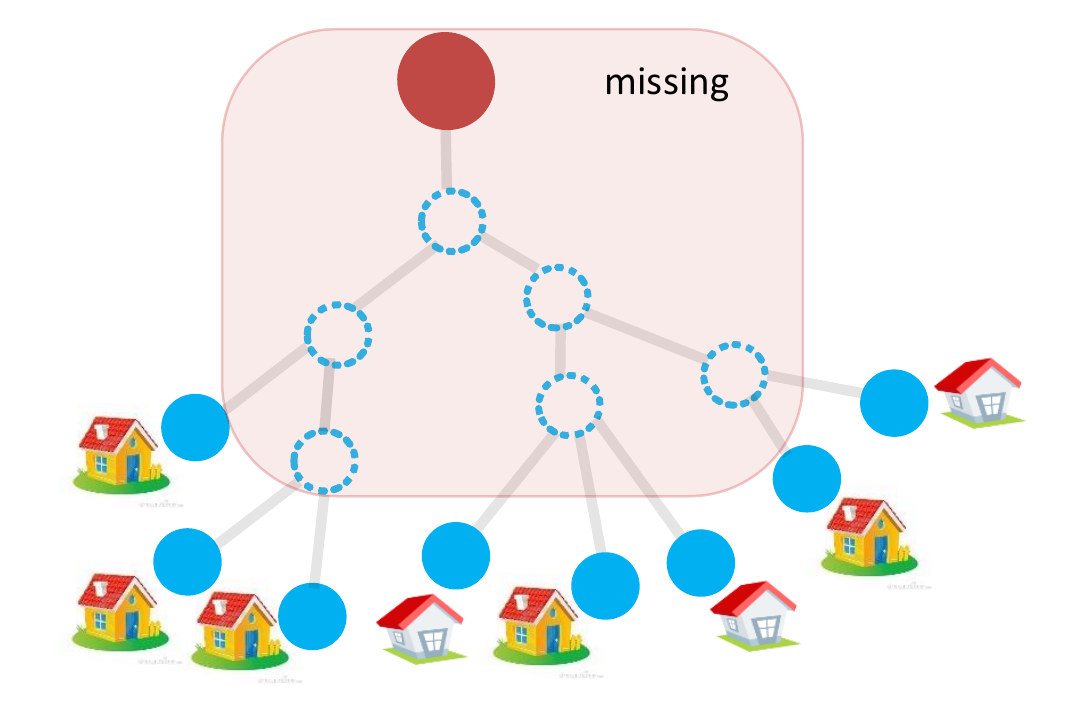}
\caption{Illustration of a radial distribution grid with a substation root node. Leaf nodes are observed and colored solid blue, while unobserved and possibly unknown intermediate nodes are denoted by dotted blue circles. All operational lines and their impedances are unknown.}
 \label{fig:radialgrid}
\squeezeup
\end{figure}
\noindent{\bf Radial Structure:}
The distribution grid is defined over graph $\mathcal G=(\mathcal V,\mathcal E)$, where the set of buses/nodes is denoted by $\mathcal V$ and the set of undirected operational lines/edges is denoted by $\mathcal E$.
The operational grid is assumed to have `radial' operational structure, that is it forms the forest consisting of disjoint trees with roots corresponding to substations.
Fig \ref{fig:radialgrid} shows a disjoint tree part of a radial grid where a red node represents substation, blue nodes represent leaf nodes and dotted nodes represent internal nodes.
Notationally, we use alphabets $a,b,c,\dots$ to represent buses/nodes and $(ab)$ stands for a line/edge between nodes $a$ and $b$. We denote $t\in\mathcal V$ as a root node (reference bus).
We denote $\mathcal P_{ab}$ as a unique path from node $a$ to node $b$ from the same (operational) tree.

\noindent{\bf Power Flow Models:}
Given a radial distribution grid on a tree structured graph $\mathcal G=(\mathcal V,\mathcal E)$, 
the grid satisfies the following Kirchhoff's law of power flow which express the complex power injection at a node via node-voltages and line-impedances as follows:
\begin{equation}\label{eq:powerflow}
p_a+iq_a=\sum_{b:(ab)\in\mathcal E}\frac{v_a^2-v_av_b\exp(i\theta_a-i\theta_b)}{z^*_{ab}}.
\end{equation}
In this equation, $z_{ab},v_a,\theta_a,p_a,q_a$ denote impedance of $(ab)\in\mathcal E$, voltage magnitude, voltage phase, active and reactive power at $a\in\mathcal V$ respectively. The substation/root/reference nodes, maintained at unit/nominal voltage, are assumed known/fixed.
Since \eqref{eq:powerflow} is non-convex, we simplify the model by making a realistic assumption that the second order terms in \eqref{eq:powerflow} is negligible.
Under this assumption, we introduce
the linearized lossless power flow equation describing the linear coupled power flow (LC-PF) model
\cite{dekasmartgridcomm,dekatcns}:
\begin{equation}\label{eq:lcpf}
\begin{split}
&p_a=\sum_{b:(ab)\in\mathcal E}\big[\beta_{ab}(\theta_a-\theta_b)+g_{ab}(v_a-v_b)\big]\\
&q_a=\sum_{b:(ab)\in\mathcal E}\big[\beta_{ab}(v_a-v_b)-g_{ab}(\theta_a-\theta_b)\big]
\end{split}
\end{equation}
where $g_{ab}=r_{ab}/(x_{ab}^2+r_{ab}^2)$, $\beta_{ab}=x_{ab}/(x_{ab}^2+r_{ab}^2)$ and $r_{ab},x_{ab}$ are resistance, reactance of line $(ab)$ respectively, i.e. $z_{ab}=r_{ab}+ix_{ab}$.
By considering only deviations from the respective steady state reference values, $p,q,v,\theta$ are modeled as random variables with zero mean (counted from the known reference values). The LC-PF model \eqref{eq:lcpf} can also be stated in the following matrix form \cite{dekatcns}
\begin{align}\label{eq:lcpf2}
&v=H^{-1}_{1/r}p+H^{-1}_{1/x}q\qquad \theta=H^{-1}_{1/x}p-H^{-1}_{1/r}q
\end{align}
where $v,\theta,p,q$ are respectively vectors of voltage magnitude, voltage phase, active and reactive power at the non-substation buses of the grid. 
$H_{1/r},H_{1/x}$ represents the reduced weight Laplacian matrices for $\mathcal G\setminus\{t\}$ where $1/r_{ab},1/x_{ab}$ are used edge-weights $(ab)$ respectively.\footnote{$\mathcal G\setminus\{t\}$ denotes a subgraph of $\mathcal G$ induced by $\mathcal V\setminus\{t\}$.}
We remind that $v_a,\theta_a,p_a,q_a$ are normalized to have zero mean. 

\section{Topology and Impedance Learning Algorithm}\label{sec:main}
In this section, we introduce our main algorithm for learning topology and impedances. We assume that time-stamped observations of voltage magnitudes, active and reactive injections at the end-nodes are available to the observer. Our algorithm is built on the notion of additive `distance' defined as a distance over the graph which thus satisfies the weighted metric property, $d(a,b)=
\sum_{(cd)\in\mathcal{P}_{ab}}d(c,d)$. We first estimate the distance, and then utilize the recursive grouping algorithm \cite{choi2011learning} to learn operational topology of the grid.
Before introducing our algorithm, let us make the following assumption about the missing intermediate nodes.
\begin{assumption}\label{asm:deg3}
All missing intermediate nodes have a degree at least 3.
\end{assumption}
We note that Assumption \ref{asm:deg3} is necessary to recover the true topology of the grid. 
See Assumption 2 in \cite{dekasmartgridcomm} for the details.
In addition, we assume that the complex power injections at different nodes are uncorrelated
\begin{assumption}\label{asm:indep}
$\mathbb{E}[p_ap_b]=\mathbb{E}[q_aq_b]=\mathbb{E}[p_aq_b]=0~~\forall a\ne b$.
\end{assumption}
As considered in prior studies \cite{bolognani2013identification, dekatcns}, Assumption \ref{asm:indep} is well-justified over sufficiently short time intervals while considering deviations of injections at end-users. Further, for intermediate nodes that involved in separation of power into downstream lines and without any major nodal usage, leakage or device losses cause the net power injection, and hence may be considered as independent from the rest. Note also that the Assumption 2 does not specify the class of distributions that can model individual node’s power injection. 
It is applicable when nodal injections are negative (loads), positive (due to local generation) or are a mixture of
both. In a future work, we will relax this assumption and discuss learning in the presence of correlated end-user injection profiles that are only uncorrelated to injections at intermediate nodes.

Now, we refer the following key property of the inverse of a reduced weight Laplacian matrix which is necessary to define the grid-based distance metric \cite{dekatcns}
\begin{equation}\label{eq:invlaplacian}
H^{-1}_{1/r}(a,b)=\sum_{(cd)\in\mathcal{P}_{at}\cap\mathcal{P}_{bt}}r_{cd}.
\end{equation}
See Section 4 in \cite{dekatcns} for details.
In \eqref{eq:invlaplacian}, $\mathcal{P}_{ab}$ is the unique path from $a$ to $b$, $r_{ab}$ is the resistance of $(ab)\in\mathcal E$ and $t$ is the root (reference bus) of the grid. One can also derive a similar formulation for line reactances $x_{ab}$ and $H^{-1}_{1/x}$. 

Under Assumption \ref{asm:indep} and using \eqref{eq:lcpf2} in the case of observed nodes $a,b$, one derives the following identity
\begin{equation}\label{eq:dist3}
\begin{split}
&\mathbb{E}[v_ap_b]=H^{-1}_{1/r}(a,b)\mathbb{E}[p_b^2]+H^{-1}_{1/x}(a,b)\mathbb{E}[p_bq_b]\\
&\mathbb{E}[v_aq_b]=H^{-1}_{1/r}(a,b)\mathbb{E}[p_bq_b]+H^{-1}_{1/x}(a,b)\mathbb{E}[q_b^2]
\end{split}
\end{equation}
where $\mathbb{E}[v_ap_b],\mathbb{E}[v_aq_b],\mathbb{E}[p_b^2],\mathbb{E}[p_bq_b],\mathbb{E}[q_b^2]$ are quantities that can be computed from measurements at observed nodes $a$ and $b$. Notice that ability to estimate the expectations in \eqref{eq:dist3} implies that one can also estimate the value of $H^{-1}_{1/r}(a,b)$ and $H^{-1}_{1/x}(a,b)$ for any observed $a,b\in\mathcal V$ unless $\mathbb{E}[p_b^2]\mathbb{E}[q_b^2]=(\mathbb{E}[p_bq_b])^2$.
To avoid such pathological situation, we make the following assumption.
\begin{assumption}\label{asm:bound}
There exists a constant $\lambda$ such that for all node $a\in\mathcal V$, $\big|\mathbb{E}[p_a^2]\mathbb{E}[q_a^2]-(\mathbb{E}[p_aq_a])^2\big|\ge\lambda.$
\end{assumption}
Using the estimated $H^{-1}_{1/r}(a,b)$, we can now estimate the resistance distance (effective resistance) between observed nodes $a,b$ as
\begin{equation}\label{eq:dist4}
d_r(a,b)=\sum_{(cd)\in\mathcal P_{ab}}r_{cd}=H^{-1}_{1/r}(a,a)+H^{-1}_{1/r}(b,b)-2H^{-1}_{1/r}(a,b)
\end{equation}
Note that the effective resistance is an additive distance metric between nodes $a$ and $b$ in the grid. Similarly, one can also estimate the additive reactance distance $d_x(a,b)$ between observed nodes $a,b$.
Once we estimate $d_r(a,b)$ for all pairs of observed nodes, we can utilize the recursive grouping algorithm (RG) \cite{choi2011learning} which directly leads us, under Assumption \ref{asm:deg3}, to consistent topology and impedance estimation of the power grid.

\begin{figure*}
\centering
 \begin{subfigure}{0.12\textwidth}
  \includegraphics[width=\textwidth]{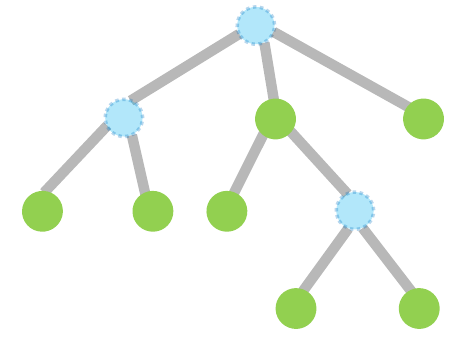}
   \caption{}
    \label{fig:rg1}
  \end{subfigure}
  ~
  \begin{subfigure}{0.12\textwidth}
    \includegraphics[width=\textwidth]{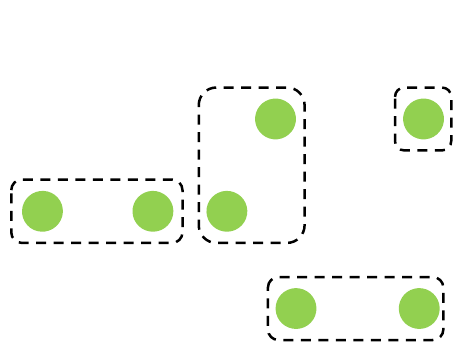}
    \caption{}
    \label{fig:rg2}
  \end{subfigure}
  ~
  \begin{subfigure}{0.12\textwidth}
    \includegraphics[width=\textwidth]{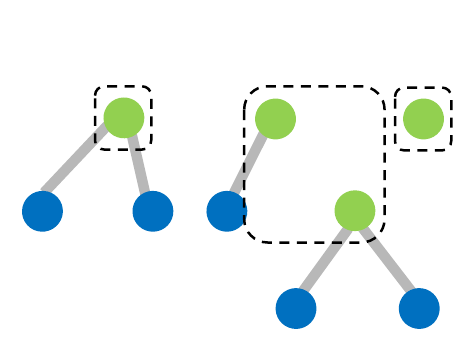}
    \caption{}
    \label{fig:rg3}
  \end{subfigure}
  ~
  \begin{subfigure}{0.12\textwidth}
    \includegraphics[width=\textwidth]{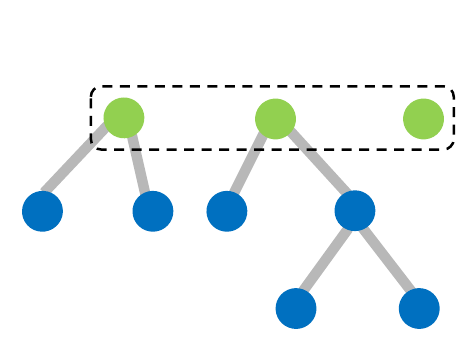}
    \caption{}
    \label{fig:rg4}
  \end{subfigure}
  ~
  \begin{subfigure}{0.12\textwidth}
    \includegraphics[width=\textwidth]{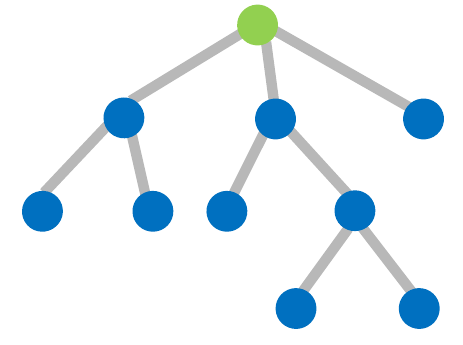}
    \caption{}
    \label{fig:rg5}
  \end{subfigure}
  ~
  \begin{subfigure}{0.12\textwidth}
    \includegraphics[width=\textwidth]{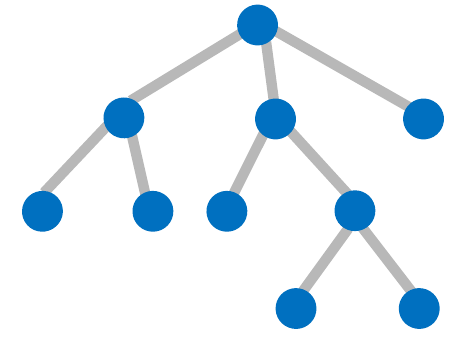}
    \caption{}
    \label{fig:rg6}
  \end{subfigure}
 \caption{(a) an original topology where skyblue nodes are missing and green nodes are observed ($\mathcal O$ in Algorithm \ref{alg:rg}) (b) a partition $\Pi$ of $\mathcal O$ generated by sibling groups and its parent in the first iteration of RG where each set in $\Pi$ is marked by the dashed line (c) updated $\mathcal O$ after the first iteration of RG and a partition $\Pi$ of $\mathcal O$ in the second iteration of RG (d) updated $\mathcal O$ after the second iteration of RG and a partition $\Pi$ of $\mathcal O$ in the third iteration of RG (e) a result after the third iteration of RG (f) a recovered topology}
 \label{fig:rg}
\end{figure*}
\subsection{Recursive Grouping Algorithm with Exact Distance}
The recursive grouping algorithm (RG) is an algorithm which recovers the true radial topology given any additive distance $d(\cdot,\cdot)$ of observed nodes on the tree where it requires to observe every leaf nodes.
Now, we first make the ideal setting assumption that exact values of $d(\cdot,\cdot)$ are known for every pair of observed nodes.
We now need to introduce the following lemma \cite{choi2011learning}.
\begin{lemma}\label{lem:dist}
For $\Phi_{abc}:=d(a,c)-d(b,c)$, the following relation holds:
\begin{itemize}
  \item[(i)] $\Phi_{abc}=d(a,b)$ for all $c\in\mathcal V\setminus\{a,b\}$ if and only if $a$ is a leaf node and $b$ is its parent.
  \item[(ii)] $-d(a,b)\le \Phi_{abc}=\Phi_{abc^\prime}\le d(a,b)$ for all $c,c^\prime\in\mathcal V\setminus\{a,b\}$ if and only if $a,b$ are leaf nodes with common parent, i.e., they belong to the same group of siblings.
\end{itemize}
\end{lemma}
Using Lemma \ref{lem:dist} (i), one can figure out the parent-child relationship for a set of observed nodes $\mathcal O$.
In addition, Lemma \ref{lem:dist} (ii) enables us to find the groups/sets of siblings of $\mathcal O$.

Now we are ready to describe how the RG works. The RG steps are illustrated in Figure \ref{fig:rg}. The input of RG is a set of observed nodes $\mathcal O\subset\mathcal V$ and the additive distance $d(a,b)$ for all $a,b\in\mathcal O$. 
For example, in Figure \ref{fig:rg1}, green nodes represent $\mathcal O$. 
First, RG finds groups of siblings and parents of a node using Lemma \ref{lem:dist}, also as illustrated in Figure \ref{fig:rg2}.
After recovering the parent-child, sibling relationships, it adds edges to all identified parent-child pairs.
For siblings without observed parent, RG adds a new node for a potential parent and adds edges to the newly added parents and its children.
The procedure is illustrated in Figure \ref{fig:rg3}.
Once nodes/edges update is done, RG updates the distance $d(\cdot,\cdot)$ between the newly added parents.
For siblings $a,b\in\mathcal O$ and their newly added parent $h$, the distance $d(a,h)$ is calculated by
\begin{equation}\label{eq:dist1}
d(a,h)=\frac{1}{2}(d(a,b)+\Phi_{abc})
\end{equation}
for any $c\in\mathcal O$.
Also, for any $c\in\mathcal O$, RG also computes $d(c,h)$ by
\begin{equation}\label{eq:dist2}
d(c,h)=d(a,c)-d(a,h).
\end{equation}
Finally, RG updates the set $\mathcal O$ with (newly added) parents and nodes which have no (established) relations which requires to update the parent-child, sibling relationships at the next iteration of RG.
For example, green nodes in Figure \ref{fig:rg3} is an updated $\mathcal O$.
After updating $\mathcal O$, RG starts over the whole procedure anew unless $|\mathcal O|\le 2$, which implies that we can add an edge to remaining vertices or only a single vertex left.
Figure \ref{fig:rg4}-\ref{fig:rg6} illustrates advanced iterations of the RG (following the first one). Formal description of RG is given in the Algorithm \ref{alg:rg}.

\begin{algorithm}\caption{Recursive Grouping Algorithm ($\mathtt{RG})$}\label{alg:rg}
\begin{algorithmic}[1]
\State {\bf Input:} $\mathcal O$, $\{d(a,b):a,b\in\mathcal O\}$
\State {\bf Output:} $(\mathcal V,\mathcal E)$, $\{d(a,b):a,b\in\mathcal V\}$
\State {\bf Initialization:} $\mathcal V=\mathcal O, \mathcal E=\emptyset$
\While{$|\mathcal O|>2$}
\State $\mathcal O_{NEW}\leftarrow\emptyset$.
\State Compute $\Phi_{abc}=d(a,c)-d(b,c)$ for all $a,b,c\in \mathcal O$.
\State Find a coarsest partition $\Pi$ of $\mathcal O$ such that any two nodes in $S\in\Pi$ are either leaves and sibling, or a parent and a leaf child.\footnotemark
\For{$S\in\Pi$}
\If{$|S|=1$}
\State $\mathcal O_{NEW}\leftarrow \mathcal O_{NEW}\cup S$.
\ElsIf{a parent $p_S\in S$ exists}
\State $\mathcal E\leftarrow\mathcal E\cup\big\{(p_Sa):a\in S\setminus\{p_S\}\big\}$
\State $\mathcal O_{NEW}\leftarrow \mathcal O_{NEW}\cup\{p_S\}$
\Else
\State Add a parent $h_S$ of $S$ as follows
\State $\mathcal V\leftarrow\mathcal V\cup\{h_S\}$
\State $\mathcal E\leftarrow\mathcal E\cup\big\{(h_Sa):a\in S\setminus\{h_S\}\big\}$
\State $\mathcal O_{NEW}\leftarrow \mathcal O_{NEW}\cup\{h_S\}$
\EndIf
\EndFor
\State Update $d(\cdot,\cdot)$ for $\mathcal O_{NEW}$ using \eqref{eq:dist1}, \eqref{eq:dist2}.
\State $\mathcal O\leftarrow \mathcal O_{NEW}$.
\EndWhile
\If{$|\mathcal O|=2$}
\State $\mathcal E\leftarrow\mathcal{E}\cup\{(ab):a,b\in\mathcal O, a\ne b\}$
\EndIf
\end{algorithmic}
\end{algorithm}
\footnotetext{$\Pi$ is a coarsest partition if for any $\Pi^\prime$ and for any $S^\prime\in\Pi^\prime$, there exists $S\in\Pi$ such that $S^\prime\subset S$. The coarsest partition $\Pi$ in Algorithm \ref{alg:rg} represents a collection of sets of siblings and their parent.}
Overall, we propose the following two stage algorithm for topology learning of grids with missing nodes for learning the topology and impedance of the grid.
\begin{itemize}
  \item[1.]For all pairs of observed nodes $a,b\in\mathcal O$, calculate ${d}_r(a,b)$ and ${d}_x(a,b)$ using \eqref{eq:dist3}, \eqref{eq:dist4} and second order moments.
  \item[2.]Recover missing nodes and lines using the recursive grouping algorithm.
\end{itemize}
The formal statement of the algorithm is presented in Algorithm \ref{alg:main}.
\begin{algorithm}\caption{Topology/Impedance Learning Algorithm with Missing Nodes}\label{alg:main}
\begin{algorithmic}[1]
\State Input: $\mathcal O$, $\{\mathbb{E}[v_ap_b],\mathbb{E}[v_aq_b],\mathbb{E}[p_a^2],\mathbb{E}[q_a^2],\mathbb{E}[p_aq_a]:a,b\in\mathcal O\}$
\State Output: $(\mathcal V,\mathcal E)$, $\{r_{ab},x_{ab}:(ab)\in\mathcal E\}$
\For{$a,b\in\mathcal O$}
\State $\begin{bmatrix}
  H_{1/r}^{-1}(a,b) \\ H_{1/x}^{-1}(a,b)
 \end{bmatrix}\leftarrow
 \begin{bmatrix}
  \mathbb{E}[p_b^2] & \mathbb{E}[p_bq_b] \\ \mathbb{E}[p_bq_b] & \mathbb{E}[q_b^2]
 \end{bmatrix}^{-1}\begin{bmatrix}
  \mathbb{E}[v_ap_b] \\ \mathbb{E}[v_aq_b]
 \end{bmatrix}$
 \vspace{0.03in}
\EndFor
\For{$a,b\in\mathcal O$}
\State $d_r(a,b)\leftarrow H_{1/r}^{-1}(a,a)+H_{1/r}^{-1}(b,b)-2H_{1/r}^{-1}(a,b)$
\State $d_x(a,b)\leftarrow H_{1/x}^{-1}(a,a)+H_{1/x}^{-1}(b,b)-2H_{1/x}^{-1}(a,b)$
\EndFor
\State $(\mathcal V,\mathcal E),\{d_r(a,b):a,b\in\mathcal V\}\leftarrow\mathtt{RG}(\mathcal O,\{d_r(a,b):a,b\in\mathcal O\})$
\For{$(ab)\in\mathcal E$}
\State $r_{ab}\leftarrow d_r(a,b)$,
 $x_{ab}\leftarrow d_x(a,b)$  where $d_x(a,b)$ is obtained using $(\mathcal V, \mathcal E)$
\EndFor
\end{algorithmic}
\end{algorithm}

\subsection{Recursive Grouping Algorithm with Samples}
In the practical scenario, we can only observe the approximated value $\widehat d_r$ of $d_r$ calculated from samples rather than the exact value.
Given finite number of samples, the variance of the distance is nonzero. 
To account for the variance, we allow some tolerance $\varepsilon$ for finding parent-child and sibling relationships.
In addition, we test the relationship of $a,b$ only using nodes which are close enough to both $a$ and $b$, i.e., nodes in $\mathcal K_{ab}$
where $\mathcal K_{ab}$ satisfies
$$\mathcal K_{ab}=\{c\in\mathcal O\setminus\{a,b\}:\widehat d_r(a,c),\widehat d_r(b,c)<\tau\}$$
for some constant $\tau$.
Let us now present rules which guide the relationships of nodes using samples.
\begin{itemize}
\item[(i)] Set $a$ as a parent of $b$ if $|\widehat d_r(a,b)-\widehat\Phi_{abc}|\le\varepsilon$ for all $c\in\mathcal K_{ab}$.
\item[(ii)] Set $a,b$ as siblings if $\underaccent{c\in \mathcal K_{ab}}{\max}\widehat\Phi_{abc}-\underaccent{c\in\mathcal K_{ab}}{\min}\widehat\Phi_{abc}\le\varepsilon.$ 
\end{itemize}
One can observe that the newly introduced rules are equivalent to the RG rules with exact $d_r(\cdot,\cdot)$ except for a tolerance $\varepsilon$. 
Update of the distance is done in a similar manner.
For $a\in\mathcal O$ and its newly added parent $h$, one sets
\begin{align*}
&\widehat d_r(a,h)=\\
&~\frac{1}{2(|\mathcal C(h)|-1)}\sum_{b\in\mathcal C(h)\setminus a}\left(\widehat d_r(a,b)+\frac{1}{|\mathcal K_{ab}|}\sum_{c\in \mathcal K_{ab}}\widehat\Phi_{abc}\right)
\end{align*}
where $\mathcal C(h)$ denotes the children set of $h$.
Likewise, update of the distance for $c\notin\mathcal C(h)$,
\begin{align*}
\widehat d_r(c,h)=\frac{1}{|\mathcal C(h)|}\sum_{a\in\mathcal C(h)}\left(\widehat d_r(a,c)-\widehat d_r(a,h)\right).
\end{align*}

\begin{figure}[t]
\centering
  \begin{subfigure}{0.24\textwidth}
    \includegraphics[width=\textwidth,height=.9\textwidth]{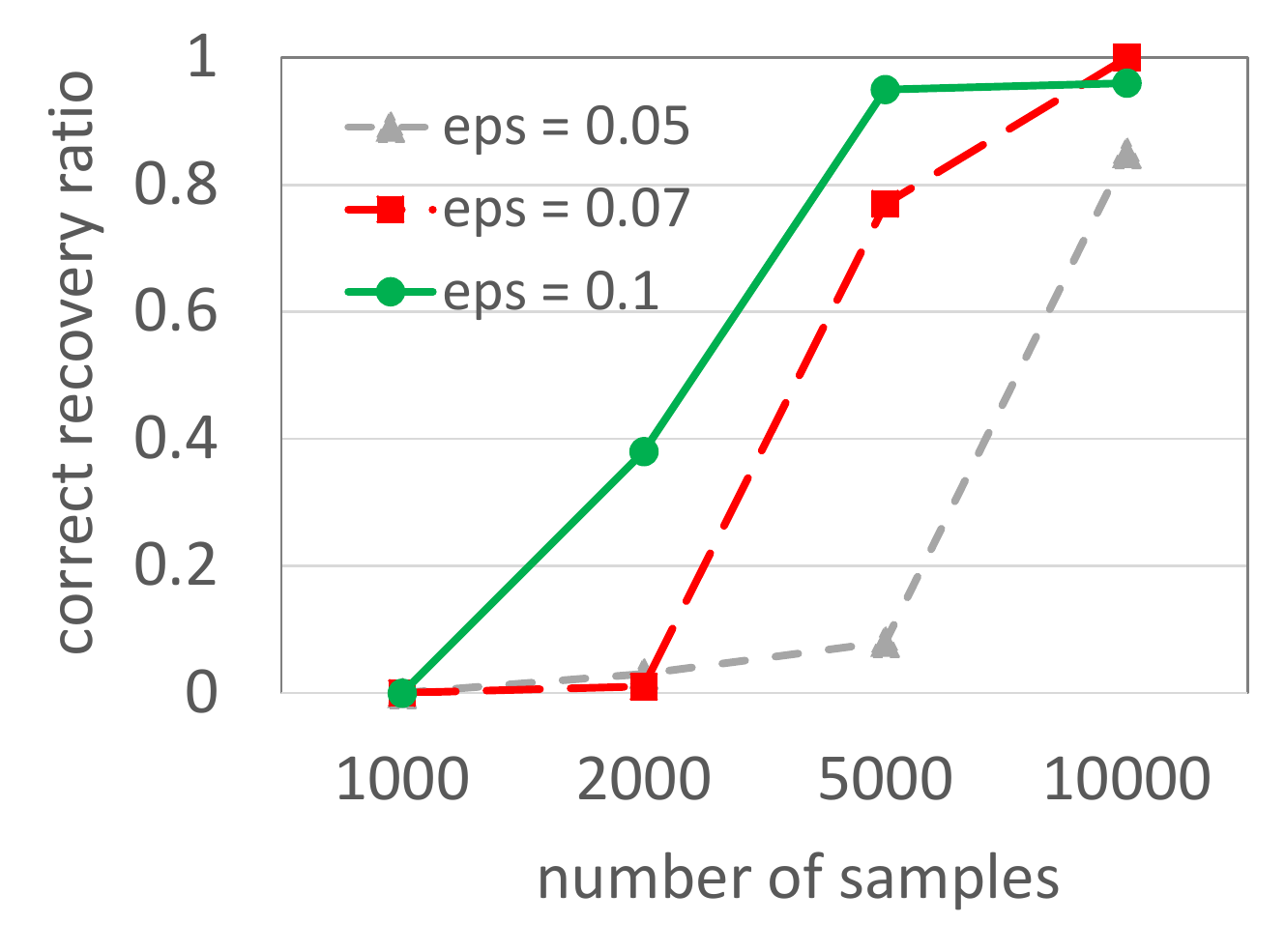}
  \end{subfigure}
  \begin{subfigure}{0.24\textwidth}
    \includegraphics[width=\textwidth,height=.87\textwidth]{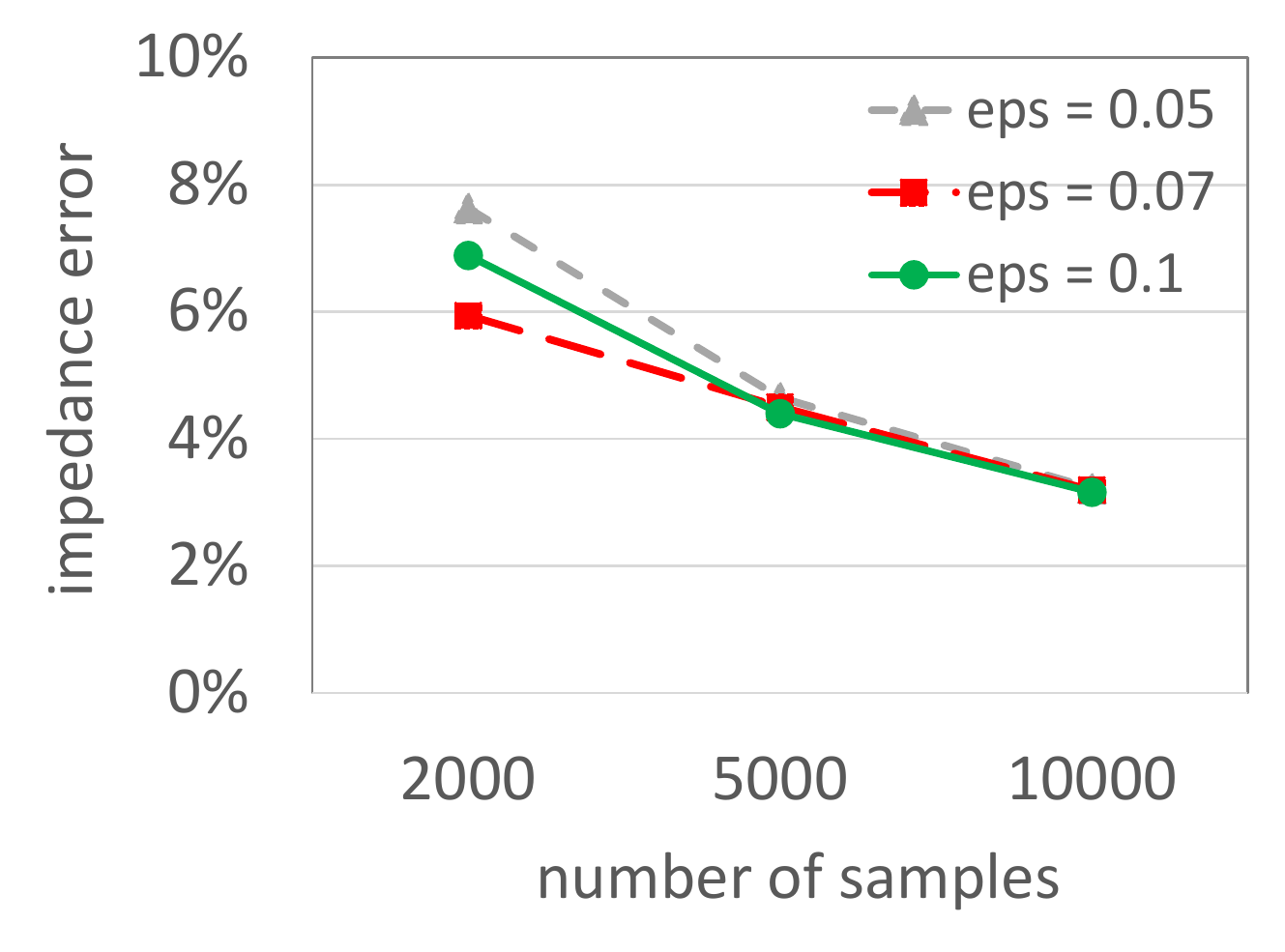}
    \label{fig:syn100}
  \end{subfigure}
\caption{Results for synthetic grids with 100 vertices averaged over 100 random radial grids. (a) Topology recovery accuracy (b) realtive impedance error. ``eps" in the graphs denotes $\varepsilon$.}  
\label{fig:syn}
\end{figure}

\textbf{Sample and Computational Complexity}: It can be shown that our algorithm terminates in $O(d|\mathcal V|^3)$ steps where $d$ is the depth of the grid. Further, our algorithm requires (under some mild assumptions) only $O(|\mathcal V|\log|\mathcal V|)$ samples to correctly recover the operational topology and line impedances. 
The exact analysis of the complexity results will be provided in the extended version.
\section{Experiments}\label{sec:experiments}
In this section, we present experimental results of our algorithm on custom and IEEE models.

\noindent{\bf Tolerance for Experiments:}
To utilize Algorithm \ref{alg:main} directly, one should carefully choose the tolerance $\varepsilon$ depending on the number of samples. While small $\varepsilon$ results in error as RG would not update the relationship, loose structure with small number of missing nodes is estimated if one chooses large $\varepsilon$. Instead in our experiments, we dynamically change $\varepsilon$ in RG, i.e. we increase $\varepsilon$ by $\varepsilon\leftarrow 1.5\varepsilon$ if no relationship is updated while we reset $\varepsilon$ after updating the relationships.

\noindent{\bf Custom Examples:}
In each of our simulation runs we construct a random tree with 100 nodes and  maximum degree 5. The line resistance and reactance are independently sampled from the distribution $\text{Unif}(0.1,0.2)$.\footnote{$\text{Unif}(a,b)$ is a uniform distribution on an interval $[a,b]$.} Complex nodal power injections are sampled from the independent normal distribution, i.e., $p_a,q_a\sim N(0,1)$. Using the generated complex power injections, we compute nodal voltage magnitudes and phases using LC-PF equation \eqref{eq:lcpf2}. The input of the algorithm is the complex power injections and voltage magnitude of the leaf/end-user nodes.

Under this setting, we run experiments changing the number of samples from 1000 to 10000, and also changing tolerance $\varepsilon$. To quantify performance of our algorithm, we record average accuracy in the recovered topology and impedances over 100 random radial grids. Figure \ref{fig:syn} shows the correct recovery ratio and the average error in estimating line impedances. The average error is defined for the grid with correct recovery according to $\frac{1}{2|\mathcal{E}|}\sum_{(ab)\in\mathcal E}\frac{|r_{ab}-\hat r_{ab}|}{|r_{ab}|}+\frac{|x_{ab}-\hat x_{ab}|}{|x_{ab}|}$. One observes that our algorithm recover line impedances with small error even in the demanding case of 1000 samples. We also observe that larger $\varepsilon$ results in a higher accuracy for the small number of samples but it becomes less accurate for the large number of samples (compare $\varepsilon=0.1$ to $\varepsilon=0.07$).

\begin{figure}[ht]
  \centering
\includegraphics[width=0.3\textwidth]{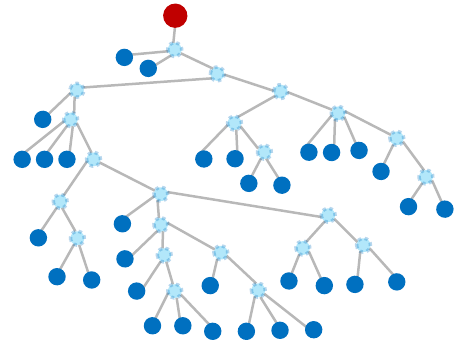}
\caption{Illustration of an IEEE radial distribution grid with 33 leaf/end-user nodes and 22 internal nodes.}
  \label{fig:ieee123grid}
\end{figure}

\begin{figure}[ht]
\centering
  \begin{subfigure}{0.35\textwidth}
    \includegraphics[width=\textwidth]{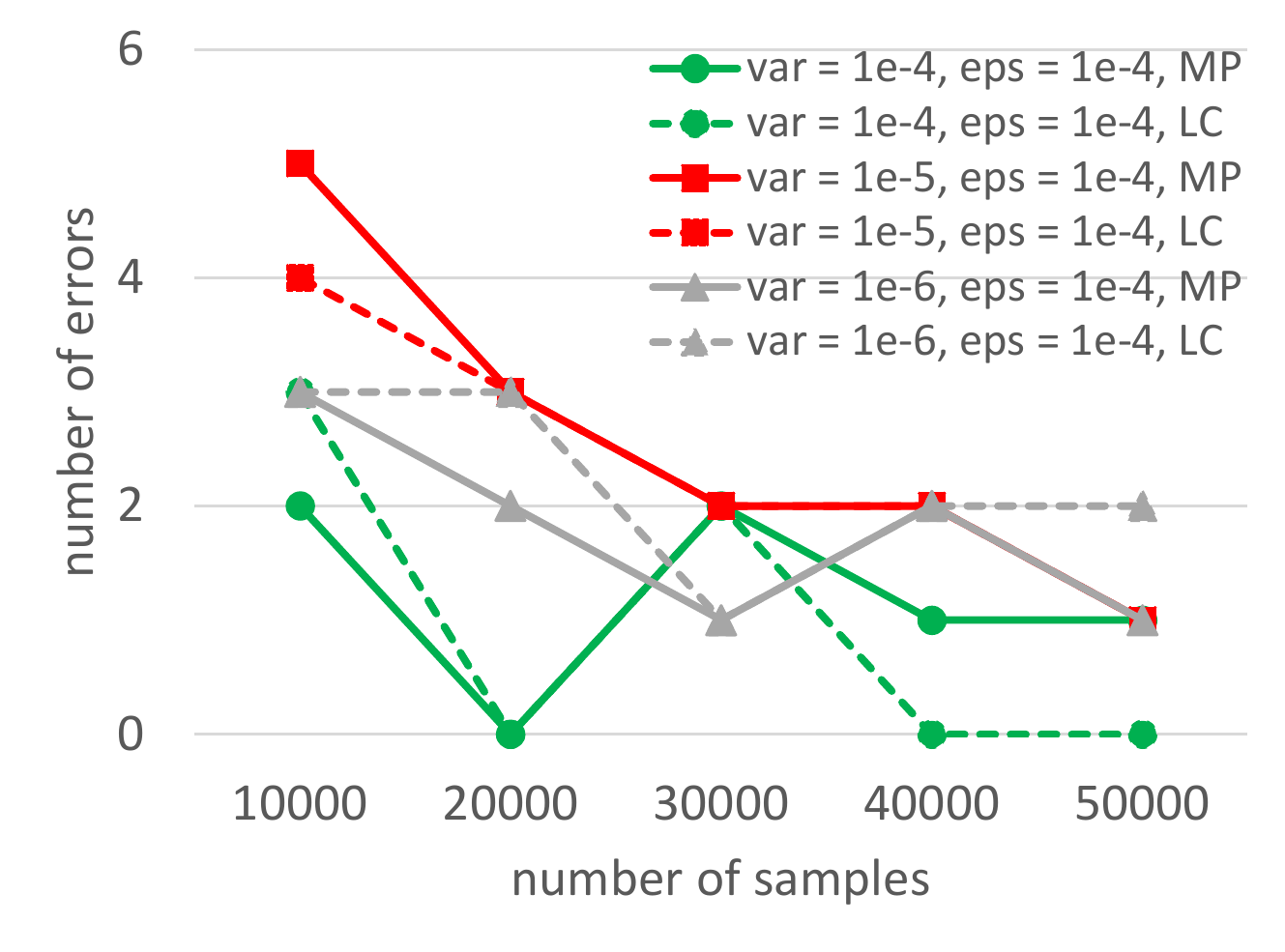}
  \end{subfigure}\hfill
  \begin{subfigure}{0.35\textwidth}
    \includegraphics[width=\textwidth]{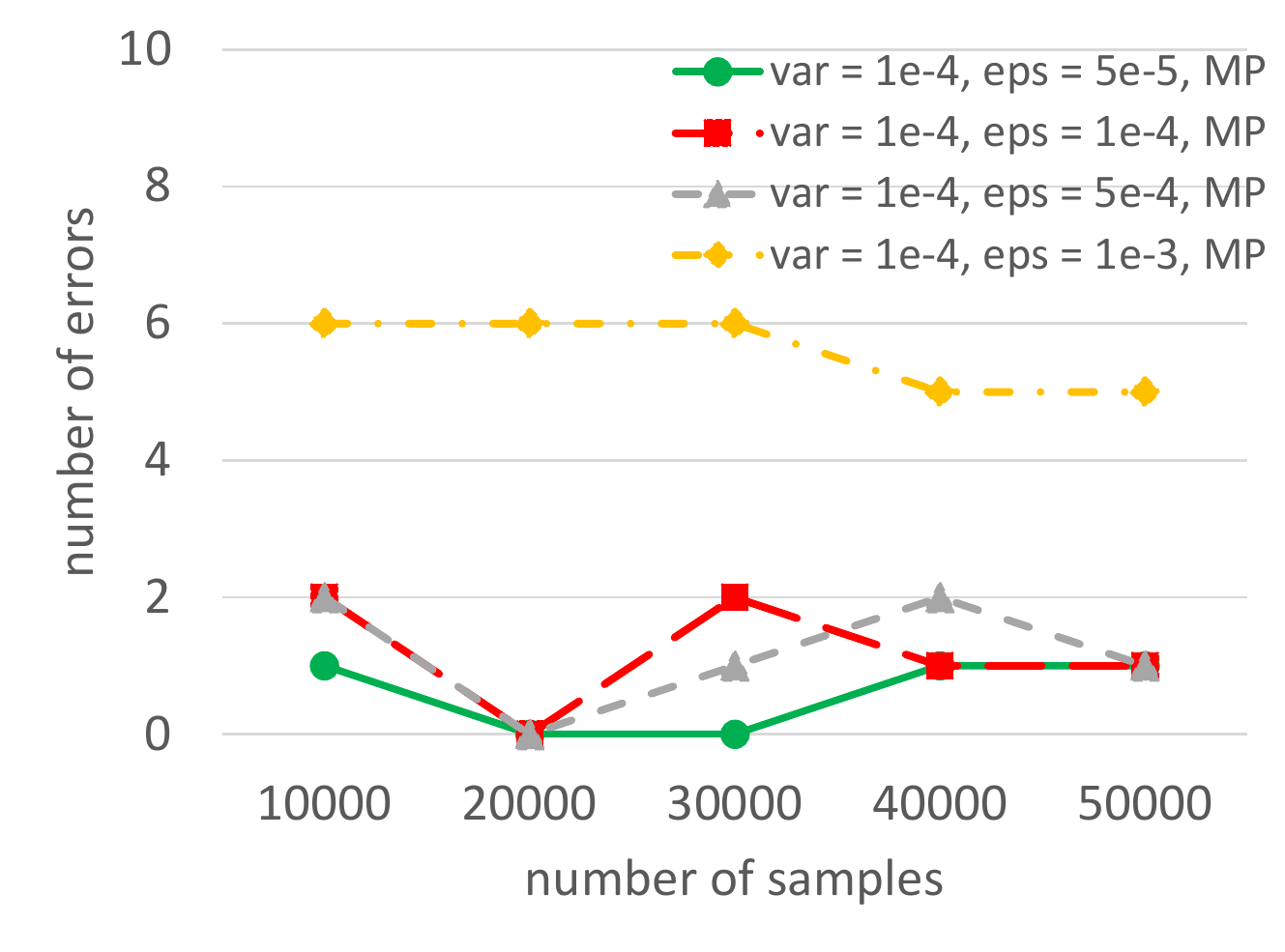}
  \end{subfigure}
\caption{Error in estimated topology observed with (a) changing variance of nodal injections (b) changing ``eps" in algorithm. ``var" denotes the variance of the complex power injections. MP, LC imply that the samples are from MATPOWER and LC-PF respectively.}  
\label{fig:ieee123}
\end{figure}

\noindent{\bf IEEE models:}
For the more realistic experiments we use a IEEE model with 56 nodes \cite{bolognani2016existence} where the topology was modified (to be radial) and the internal nodes all to be of degree $\ge 3$. The modified grid is illustrated in Figure \ref{fig:ieee123grid}.
We generate the complex power injections from the independent normal distribution as in the case of the custom models.
From the complex power injections, we obtain the corresponding voltage magnitude by using ac power flow solver in MATPOWER \cite{zimmerman2011matpower}. Under this setting, we measure performance of our learning algorithm by varying the input number of samples, the variance of the complex power injection and the tolerance value, $\varepsilon$. In particular, as a way to quantify errors, we count the number of edge difference between the recovered topology and the true topology. We also compare the performance of our algorithm with MATPOWER samples and LC-PF samples generated with the same complex power injections. Figure \ref{fig:ieee123} shows our IEEE model experimental results. Observe that the algorithm works similarly for both MATPOWER samples and the LC-PF samples. Similar to what we saw in the custom model experiments, the algorithm performance decreases as the tolerance $\varepsilon$ increases.
\section{Conclusion}\label{sec:conclusions}
Topology learning of the distribution grids in real time from sparse data is critically important for a number of operational/control applications.
In this manuscript, we propose a novel algorithm which recovers topology and line impedances by only using measurements at the end-user nodes.
In this approach we utilize LC-PF model to approximate the resistance distance (also called effective resistance) between any two observed nodes and apply the recursive grouping algorithm to recover the topology. Furthermore, our experimental results, derived for custom (randomized) and IEEE models, shows that the algorithm performs remarkably well. In the future, we plan to extend our algorithm to the case of correlated injections/consumptions and also attempt to generalize to the case of sparse but loopy operations grids/graphs.


\bibliography{reference,sigproc,FIDVR,SmartGrid,voltage,trees}

\end{document}